**Data Standards in Audiology: A Mixed-Methods Exploration of Community Perspectives and Implementation Considerations**

Charlotte Vercammen [1,2,3], Antje Heinrich [2,4,#], Christophe Lesimple [1,#], Alessia Paglialonga [5,#], Jan-Willem A. Wasmann [6,#], & Mareike Buhl [7,*]

[1] Sonova AG, Research & Development, Stäfa, Switzerland

[2] Manchester Centre for Audiology and Deafness, School of Health Sciences, Faculty of Biology, Medicine and Health, University of Manchester, Manchester, United Kingdom

[3] Department of Neurosciences, Research Group Experimental Oto-Rhino-Laryngology, KU Leuven—University of Leuven, Belgium

[4] NIHR Manchester Biomedical Research Centre, Manchester, United Kingdom

[5] Cnr-Istituto di Elettronica e di Ingegneria dell'Informazione e delle Telecomunicazioni (CNR-IEIIT), Milan, Italy

[6] Department of Otorhinolaryngology, Donders Institute for Brain, Cognition and Behavior, Radboud university medical center, Nijmegen, The Netherlands

[7] Université Paris Cité, Institut Pasteur, AP-HP, INSERM, CNRS, Fondation Pour l'Audition, Institut de l'Audition, IHU reConnect, F-75012 Paris, France

[#]These authors contributed equally (listed in alphabetical order).

* Corresponding author: Mareike Buhl

Email: mareike.buhl@pasteur.fr

**Abstract**

***Objective:*** This study addresses conceptual issues around data standardisation in audiology, and outlines steps toward achieving it. It reports a survey of the computational audiology community on their current understanding, needs, and preferences concerning data standards. Based on survey findings and a panel discussion, recommendations are made concerning moving forward with standardisation in audiology.

***Design:*** Mixed-methods: 1) review of existing standardisation efforts; 2) a survey of the computational audiology community; 3) expert panel discussion in a dedicated session at the 2024 Virtual Conference of Computational Audiology.

***Sample:*** Survey: 82 members of the global community; Panel discussion: five experts.

***Results:*** A prerequisite for any global audiology database are agreed data standards. Although many are familiar with the general idea, few know of existing initiatives, or have actively participated in them. Ninety percent of respondents expressed willingness to follow or contribute to standardisation efforts. The panel discussed relevant initiatives (e.g. OMOP, openEHR, Noah) and explored both challenges (around harmonisation) and opportunities (alignment with other medical fields and conversion among approaches).

***Conclusions:*** Combining conceptual discussion with stakeholder views, the study offers guidance for implementing interoperable data standards in audiology. It highlights community support, key issues to address, and suggests paths for future work.

## Introduction

Audiological data are collected in many settings, such as hospitals and research facilities, and are increasingly stored in digital formats. The data cover diverse aspects of hearing health, including patients' hearing abilities captured by behavioural responses and objective measures, medical imaging results, anamnesis information, general health data, hearing device (HD) settings, and other sensor readings that are part of hearing or other wearable devices, web platforms, and smartphone applications. Various types of data sets exist: First, clinical data, such as routinely collected patient care data available in electronic health records (EHRs). Examples include the audiological dataset from the US Department of Veteran Affairs (Dillard et al., 2020), the Hearing Examination in Southern Denmark (HESD) database from hospitals in Southern Denmark (Cantuaria et al., 2021), multimodal EHR data on hearing, cognition, and comorbidities (Callejon-Leblic et al., 2024), and the German Cochlear Implant Registry (Stöver et al., 2023). Second, research data are increasingly published in online repositories. Examples include the Better hEAring Rehabilitation (BEAR) dataset (Sanchez-Lopez et al., 2021), the Oldenburg Hearing Health Record (OHHR) (Jafri et al., 2025), and a public health policy-making platform for hearing loss (EVOTION) (Christensen et al., 2019). These experimental datasets often include variables not routinely collected in clinical settings, such as loudness measures, binaural and temporal-spectral auditory processing outcomes, cognitive test results, questionnaire responses, HD usage data, acoustic environmental data, and smartphone application data. Third, data collected through organisations, including industrial (e.g., Hearing Instrument Manufacturers' Software Association (HIMSA; HIMSA, n.d.)) and governmental institutes (e.g., the National Institute for Occupational Safety and Health (NIOSH); Masterson et al., 2023).

Successfully combining datasets across data sources offers exciting opportunities for personalized care, remote service delivery, digital health coaches, digital clinical trials, and digital twin technology (Acosta et al., 2022). For audiology, understanding gained from combined digital data could revolutionise the delivery of hearing care. This is particularly pressing now given that the number of patients with hearing loss is increasing while the availability of hearing care by qualified professionals is decreasing (Windmill & Freeman, 2013; Wasmann et al., 2021). Combined datasets could lead to new insights into hearing loss patterns, prevalence, and relationships between hearing loss etiology and management strategies. Some examples of such research already exist and include profiling approaches by Saak et al. (2022), Sanchez-Lopez et al. (2020), and work towards a clinical decision support system as proposed by Buhl (2022). An important prerequisite for successfully and efficiently combining datasets from different sources, in addition to putting the legal aspects of data sharing in place, is adherence to a standardised data format (or especially interoperability, see below). One way to encourage compliance with and widespread adoption of data standards by the global audiological community is to put their voice at the centre of development.

### *Data standards and interoperability*

A call for the development of interoperable, community-agreed data standards of hearing health data was published by HD manufacturers nearly a decade ago (Laplante-Lévesque et al., 2016). Yet, to date, such a standard has not been achieved. One of the underlying reasons for this may be the absence of a recognised body bringing together all audiology stakeholders. After all, according to the International Organization for Standardisation (ISO), a standard is "established by consensus and approved by a recognised body, that provides […] rules, guidelines or characteristics for activities or their results", achieving "the optimum degree of order in a given context" (Benson & Grieve, 2021, Chapter 22). One way to achieve this is by adopting information models, which describe database contents in a commonly shared, unambiguous manner, including metadata specifying conditions under which the data were collected. If agreement on an information model to use is obtained, a data standard is achieved.

The audiological community collects, analyses, and interprets data, and develops and applies analysis tools (see Figure 1 for a schematic overview). Interoperability is essential to be able to combine data from different sources and tools, and data standards are the basis on which this interoperability is achieved. Interoperability is defined as "the ability of data or tools from non-cooperating resources to integrate or work together with minimal effort" (Wilkinson et al., 2016). The FAIR principles (Findable, Accessible, Interoperable, Reusable; Wilkinson et al., 2016) describe essential properties for creating machine-actionable databases, emphasising that interoperability requires "a formal, accessible, shared, and broadly applicable language for knowledge representation", while reusability necessitates "(meta)data meeting domain-relevant community standards". Particularly, semantic interoperability assures common understanding and meaning of the data described by data models and terminologies (HIMSS, n.d.). After defining the data, metadata, and formats of a database, a data standard can be applied to transform the data into a standardised format. Data standards thus facilitate data harmonisation, sharing, and analysis (Benson & Grieve, 2021, Chapter 3). In an ideal world, data standards are adopted by all stakeholders to ensure they benefit everyone.

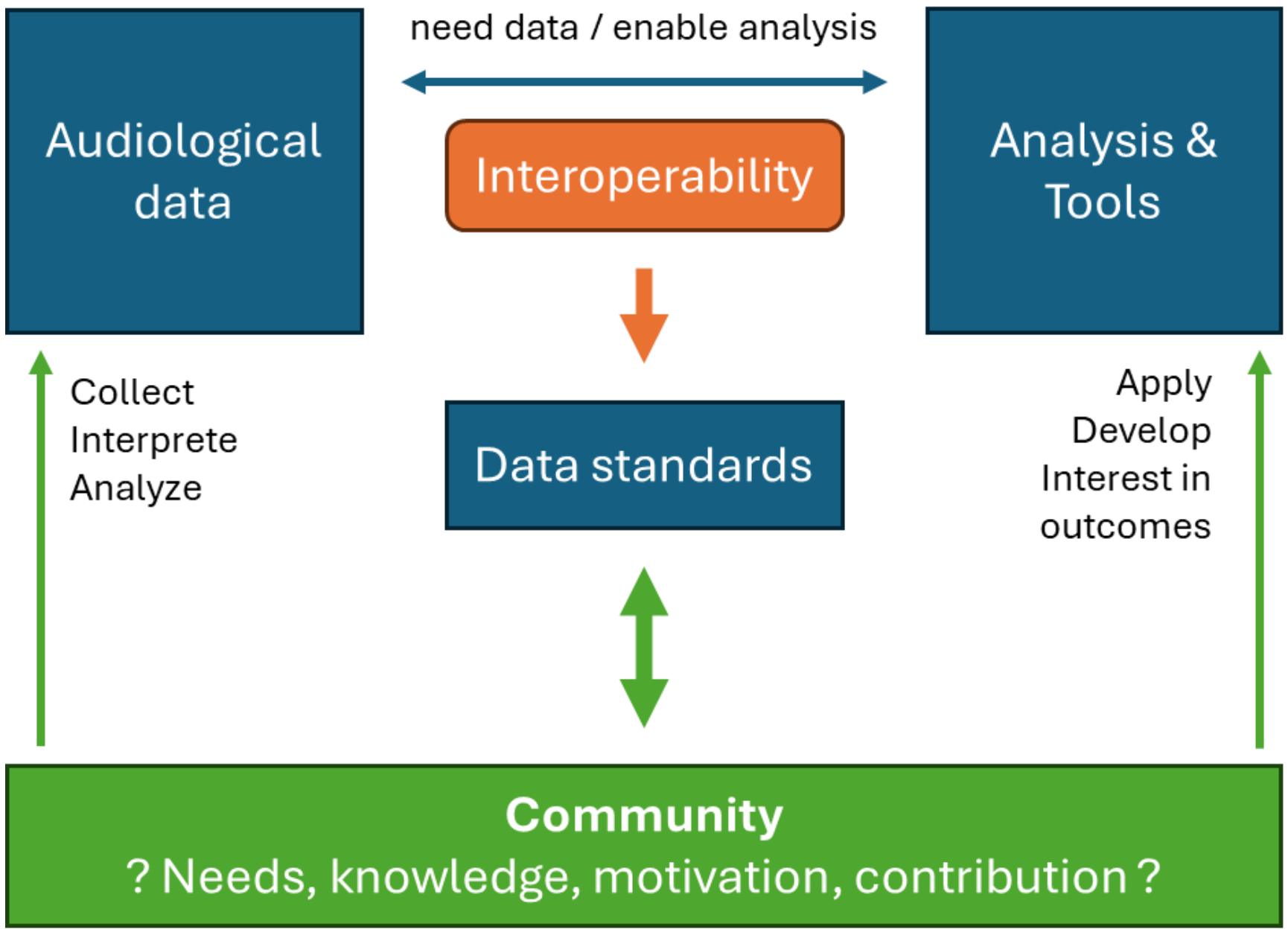

*Figure 1: Context of data standards and the role of the audiological community.*

*Application of existing (medical) data standards in audiology*

Other medical fields have been successful in establishing and adopting data standards internationally (e.g., Wulff et al., 2021; Reinecke et al., 2021; Byun et al., 2022). Consequently, their standards and methods may prove valuable starting points for the field of audiology. In addition, audiology-specific initiatives should be considered. While the list of existing initiatives may not be exhaustive, three data standard initiatives are of particular interest and will be discussed in more detail below: (1) openEHR (Beale, 2002; Haarbrandt et al., 2018); (2) the Observational Medical Outcomes Partnership (OMOP) by the Observational Health Data Sciences and Informatics (OHDSI) community (OHDSI, 2021); and (3) the Noah standard (HIMSA, n.d.). The first two initiatives are not audiology-specific, while the third is.

Data standards are developed for different purposes. In the case of the three standards considered here, the purposes are: Interoperability between electronic health records for openEHR, harmonising data for research and analysis purposes for OMOP, and ensuring clinical interoperability between audiological software and measuring equipment for HIMSA's Noah standard (see Table 1 for a detailed comparison). All three approaches have reached some maturity in audiology. For instance, the European Federation of Audiological Societies (EFAS) working group is developing openEHR audiological standards through archetype modelling and community review (EFAS Working Group, n.d.). Mehta et al. (2024) utilised the OMOP data model to build a hearing-health research database in the UK. HIMSA's Noah standard is broadly adopted in audiological clinical practice (HIMSA, n.d.) and ensures that data is logged in a common format independent of the brand of measuring software / device used by clinicians. While these three data standards are the focus of the current paper, the principles discussed could also be applied to other standards, such as the HL7 FHIR (Health Level 7 Fast Healthcare Interoperability Resources), which focuses on exchange of health data instead of organising data repositories (HL7, n.d.).

*Table 1. Comparison of main characteristics of openEHR, OMOP, and HIMSA's Noah standard (Adapted from HIMSA [n.d.], OHDSI [2021], Haarbrandt et al. [2018]).*

| **Characteristic** | **openEHR** | **OMOP** | **HIMSA's Noah Standard** |
|---|---|---|---|
| Purpose | Semantic interoperability between EHRs | Harmonizing heterogeneous data for research/analysis | Clinical interoperability for audiology software/equipment |
| Domain | Health data (not specific to a particular medical field) | Health data (not specific to a particular medical field) | Audiology |
| Data Model | Two-level methodology with reference model (basic structures, data types, functions) and archetype model (domain specific contents in archetypes and templates) | Common Data Model (CDM) based on medical terminologies | Proprietary integration framework for hearing-care software |
| Key Concept | Archetypes/templates (assuring semantic interoperability)[1] | Standardised concepts (unique identifiers) | Standardised audiological test, fitting, client record definitions |
| Community Approach | Clinical Knowledge Manager (CKM) platform[2]; open community review | OHDSI community collaboration | Managed by HIMSA consortium; limited external input |
| Tools | Open-source tools for archetype/template modelling[3] | ATHENA platform for vocabularies/concepts[9], proposition to SNOMED CT[10] , open-source tools[11] | Noah software, fitting and measurement modules |
| ETL Process* Complexity | Moderate; guided by archetypes/templates[4] | High; detailed scripting/mapping[12] | Low; direct standardised integration |
| Governance | openEHR governance board[5] (community-supported) | Community-driven (OHDSI, open science) | Consortium governance; industry-maintained |
| Use Cases | Clinical interoperability, patient care, research | Observational/population research, analytics | Clinical workflow tasks (assessment, device fitting, clinic management) |
| Geographic Adoption | Germany, Norway, Australia, Brazil, China[6], Switzerland, Spain, Finland, Sweden, …[7] | U.S. predominant, expanding internationally[13] | Globally extensive (~130,000 installations in >150 countries) |
| Application in Audiology | EFAS working group's archetype/template development[8] | U.K. research database applications[14] | Standard clinical audiology/hearing care software integration worldwide |

Note: References in table (weblinks last accessed May 7, 2025): 1 Beale (2002); 2 https://ckm.openehr.org/ckm/; 3 https://tools.openehr.org/designer/#/; 4 Haarbrandt et al., 2016; 5 https://openehr.org/governance/; 6 Marco-Ruiz et al. (2019); 7 https://openehr.org/affiliates/; 8 EFAS Working Group, (n.d); 9 https://athena.ohdsi.org/; 10 https://www.snomed.org/members/united-kingdom; 11 https://www.ohdsi.org/software-tools/; 12 Kohler et al. (2023); 13 https://www.ohdsi.org/wp-content/uploads/2024/10/OurJourney2024.pdf; 14 Mehta et al. (2024). *ETL (Extract, Transform, Load) processes define the transformation of data sets to a standardised format (Haarbrandt et al., 2016; Kohler et al., 2023), which is important for secondary use of data.

*Study objectives*

As stated above, including the intended target community in the development of data standards can greatly facilitate their adoption. Yet, it has not always been a priority. In fact, various data standardisation approaches, including those used in audiology, were developed for a specific purpose and not necessarily with community involvement. The goal of this study was to engage the relevant professional community from the beginning of the development process. It aimed to do this by engaging the general audiology and hearing sciences community, as well as the computational audiology community in particular, to 1) document their current knowledge and views, and 2) explore needs and preferences. To achieve these goals, we applied a mixed-methods approach and collected both quantitative and qualitative data in a structured online survey. This paper presents the results of this exploration and develops recommendations based on the results. Preliminary data were analysed in advance of the 2024 Virtual Conference of Computational Audiology (VCCA2024[1]), where they were then discussed with experts during a special session on “Big Data and Data Standards in Audiology”.

## Materials and Methods

*Survey*

An online survey was designed to query members of the audiology community about the following topics: (1) audiological data, i.e., how respondents worked with audiological data at present, how they stored the data, and what potential difficulties, if any, they encountered; (2) views on large, big databases in terms of opportunities, beneficiaries, and potential difficulties; (3) familiarity with and needs related to data standardisation. In addition, (4) basic demographic data were collected from survey respondents. Questions varied in whether they only allowed for one response or multiple responses. Normally, questions were followed by an open text box in which respondents could add extra explanations when they had chosen the “other” category. For a full list of survey questions, see Supplementary Material.

The survey was distributed using Google Forms. Participants were invited to take part in the survey through social media channels aimed at the general audiology and hearing science community and the attendee list of the VCCA2024. Data were collected anonymously between May and September 2024. No ethics statement was required for research on anonymous data, which does not need to comply with protection of human subjects’ regulations, and for which the Institut Pasteur Paris is legally responsible. A letter of exemption was obtained from Institut Pasteur.

*Analysis*

Exploratory data analysis of quantitative data was performed using R (R Core Team, 2024). Results for questions with categorical or ordinal answers were expressed as counts and percentages. The summed number of responses were divided by the total number of respondents to normalize the results. Qualitative data were analysed thematically by one author (M.B.).

[1] https://computationalaudiology.com/events/vcca2024/, last accessed 12 October 2025

*VCCA2024 special session*

A special session on “Big data and data standards for audiology” was organised at the VCCA in June 2024. The virtual session was free to attend following registration. The aims were to assess and present the current state of data standards in audiology, and gain insights from the community and panel members. After an introduction and motivation for the session, the preliminary survey results were presented. The remaining 35 minutes were dedicated to a panel discussion. The panel members had been invited for their experience and work relevant to the topic of data standards, e.g., with analysing data from EHRs, building a clinical database for research purposes, contributing to a data standardisation approach, statistical expertise, insights into industrial data collection, experience with other medical fields, and artificial intelligence (AI) applications to real-world clinical data. The experts were also invited to participate in the survey, to contribute their perspectives and prepare for the panel discussion. Slides containing the survey results were shared with the panel members before the session.

A list of questions to guide the panel discussion was prepared by the session chairs. These questions included: motivation for participating in the session; the panel members’ views on the survey results, e.g., whether they found any aspects striking or surprising; specific questions related to the survey results; the status of data standards in other medical fields; key stakeholders to involve; and some questions proposed by survey participants, for example, how data standards can be implemented without limiting clinical work, or questions related to infrastructure and data sharing.

The panel discussion took place as part of a virtual conference via Zoom, and the session was recorded. The qualitative analysis of the discussion was based on the transcript generated by the automatic subtitle transcript of the session.

**Results**

*Demographic characteristics of participants*

A total of 84 respondents participated in the survey; 82 respondents consented to have their responses used in a scientific report. An initial set of 74 responses was collected between May and June 2024 and discussed during the VCCA2024. The survey was kept open for an additional three months after the conference, which resulted in an additional ten responses. The full dataset, the results of which are presented below, comprised 82 respondents.

More than 50% of respondents were originally from Europe, but the survey also reached participants in — in descending order — Asia, North America, Australia, Africa, and South America. Respondents were mostly employed as researchers (65%), audiologists (41%), data scientists (10%), and educators (10%). In terms of work experience, respondents were evenly split between those with less than ten years and those with more than ten years of experience. Allowing for more than one answer, ninety-four percent indicated attending the conference for professional reasons, and 10% for private reasons — see Table 2 for complete details on the demographics of all survey respondents.

*Table 2. Demographic data of survey respondents: geographical origin, profession, years of work experience, and reasons for attending the conference. Percentage values were computed by normalizing to the total number of respondents (n=82).*

| **Where are you originally from?**<br>[One answer possible.] | | |
|---|---|---|
| | Count | Percentage |
| Europe | 44 | 54% |
| Asia | 17 | 21% |
| North-America | 9 | 11% |
| Australia | 5 | 6% |
| Africa | 4 | 5% |
| South-America | 3 | 4% |
| **What is your current profession?**<br>[Multiple answers possible.] | | |
| | Count | Percentage |
| Researcher | 53 | 65% |
| Audiologist | 34 | 41% |
| Data scientist | 8 | 10% |
| Educator | 8 | 10% |
| ENT doctor | 5 | 6% |
| Hearing aid acoustician | 1 | 1% |
| Informatician | 1 | 1% |
| I prefer not to answer | 1 | 1% |
| Other (please specify) | 7 | 9% |
| **How many years of professional experience do you have in the domain of audiology / hearing?**<br>[One answer possible.] | | |
| Experience (years) | Count | Percentage |
| 0-1 | 10 | 12% |
| 2-3 | 8 | 10% |
| 4-5 | 5 | 6% |
| 6-10 | 17 | 21% |
| 11-20 | 15 | 18% |
| 21-30 | 17 | 21% |
| > 31 | 9 | 11% |
| NA | 1 | 1% |
| **Do you attend this conference for private / professional reasons?** [Multiple answers possible.] | | |
| | Count | Percentage |
| Professional | 77 | 94% |
| Private | 8 | 10% |
| I prefer not to answer | 1 | 1% |
| Probably not attending | 1 | 1% |

*Survey section 1: Audiological data*

Figure 2A shows audiological data considered important for the field of audiology by the survey respondents, who could choose as many types of data as they thought were appropriate, from a predefined list of options. Results revealed, in descending order, audiological test data (aided and unaided); HD data logging and fitting data; benefit and satisfaction outcomes; demographic and anamnesis data; hearing screening data; data related to consumer electronics, and other health-related data. Respondents who chose other data added the following suggestions: data related to the profession of audiology (e.g., education of audiologists, how the profession is regulated in different countries, and number of available audiologists), observations from family members or caregivers, and other aspects of broader (hearing) health (e.g., information on social activities).

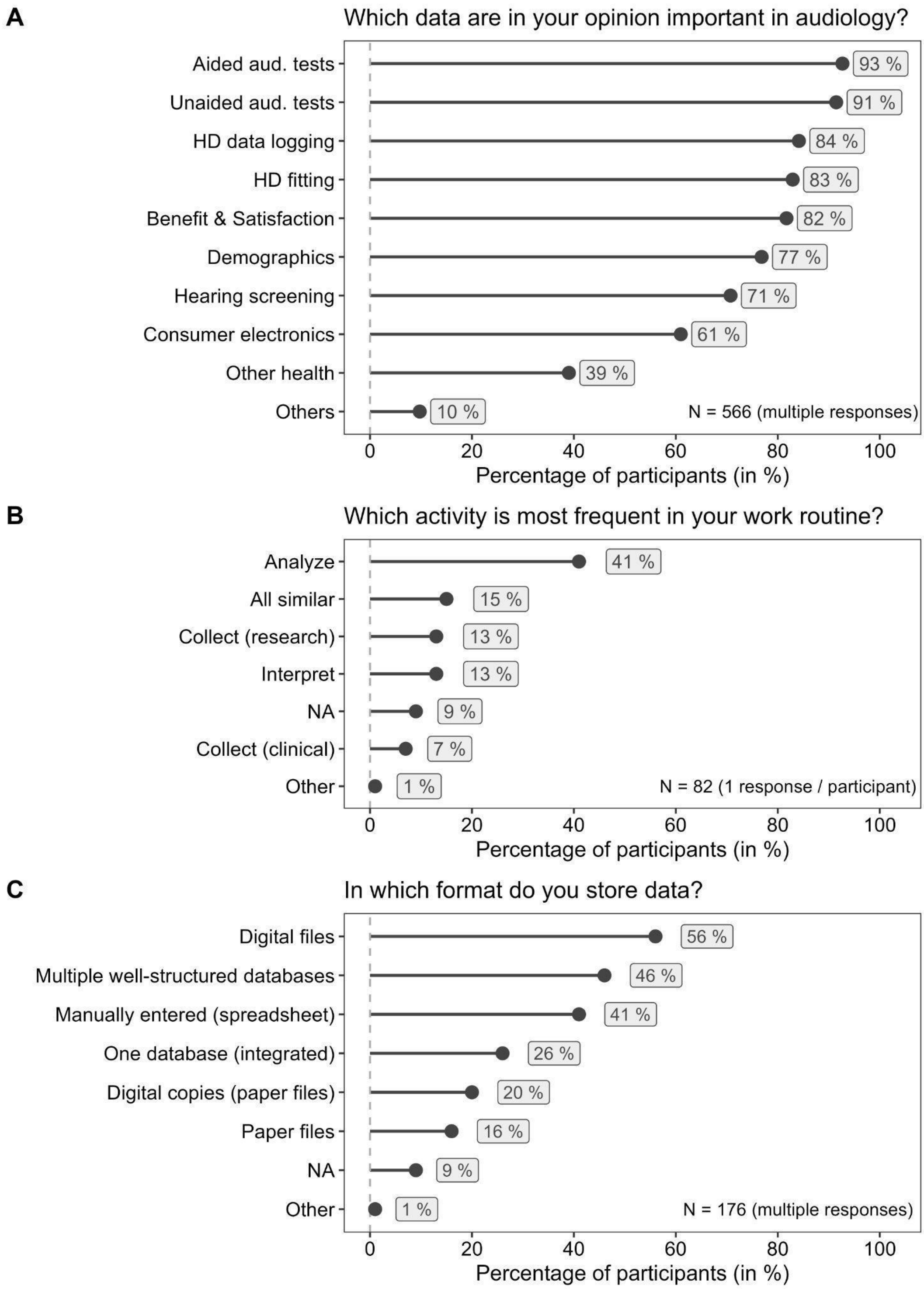

*Figure 2. Percentage of participants (x-axis) who indicated: (A) Important audiological (aud.) data categories; (B) Most common work activity with audiological data; (C) Formats in which participants stored their data. Responses were normalized to the total number of respondents (n=82).*

In response to whether they worked with audiological data, 9 out of 82 respondents (11%) indicated that they did not work with data, whereas the remaining respondents indicated that they worked with audiological data for the following tasks, in descending order (multiple responses possible): analysing data across groups of patients or participants (72%; 59 responses); collecting data as part of research studies (71%; 58 responses), interpreting data for individual patients or participants (48%; 39 responses), and collecting data as part of clinical practice (26%; 21 responses). The results of the follow-up question on which of these activities were most frequent in the work of the respondents are shown in Figure 2B. Here, the respondents could only choose one answer; 41% (34 responses) indicated they most often analysed data across patient or participant groups.

Figure 2C visualizes the formats in which survey respondents stored their data, as selected from a predefined list of options, in descending order: digital files derived from different measurement devices; multiple well-structured databases; a digital spreadsheet in which data was manually entered; one global well-structured database that integrated data collected through different devices; digital copies of paper files; and paper files.

When asked if they encountered difficulties when using data, 20% of respondents answered "no" (16 responses). More than half of respondents (52%; 43 responses) indicated encountering infrastructure problems, such as related to the use of multiple databases. In addition, 46% (38 responses) highlighted legal problems, related to matters such as consent and data sharing. A further 27% (22 responses) indicated problems related to statistics and data analysis, and 15% (12 responses) mentioned problems related to the availability of resources to analyse data. Of those respondents who had indicated "other" (9%; 7 responses), some added open text statements elaborating on the difficulties experienced, such as those related to server storage, quality of data, data cleaning, data management policies, and data types / files / procedures.

*Survey section 2: Global databases*

Figure 3A shows potential benefits envisioned by survey respondents if audiological data were available in global databases. Note that the purpose of this survey section was to gain insights into how data could be utilized if they were combined into one large database. Data standardisation would be an important prerequisite for this. Survey respondents were asked to select the two most important options, in their view, from a predefined list of categories. The most selected option was advancing research, followed by advancing patient care. Other potential benefits, in descending order, were training AI models, conducting epidemiological / population-representative studies, having access to data of parts of the world that are more difficult to reach currently, preventative health care, and other benefits. Those who had indicated "other" suggested analyses related to outcomes with cochlear implants specifically, and research & development applications.

When asked about potential beneficiaries of large, global databases (Figure 3B), patients and researchers were the most chosen options, followed by, in descending order, clinicians, health policy-makers, health insurers, and other beneficiaries (specified by respondents as manufacturers of hearing technology).

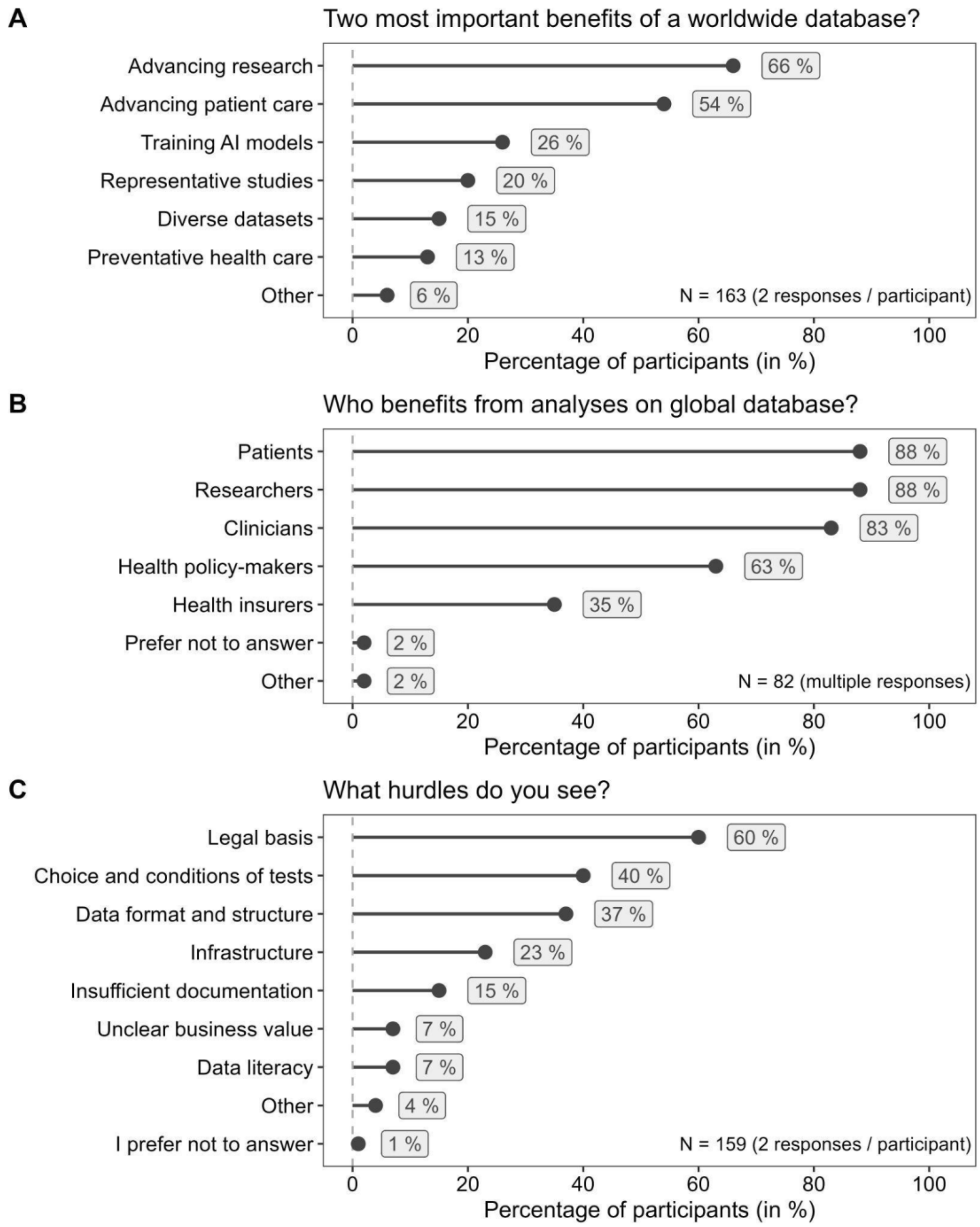

*Figure 3. Percentage of participants (x-axis) who indicated: (A) Potential benefits from analyses on combined global datasets; (B) Who would benefit from such analyses; and (C) What could hinder such analyses. Participants were asked to select the two most important options, in their view, for panels (A) and (C), while multiple responses were possible for panel (B). Responses were normalized to the total number of respondents (n=82).*

In addition to the two questions about potential benefits and beneficiaries, respondents were encouraged to “dream big” and describe in their own words what they would do with “big audiological data”, i.e., if all audiological data worldwide would be accessible in one database, in a common format. The resulting 41 open text responses were manually grouped into themes by one author (M.B.), with a theme clustering a minimum of at least five responses. Five themes emerged from the analysis (see Table 3): predictors of outcomes one could investigate using such a large database; hearing loss prevention and awareness of hearing loss; accompanying patients on their personal hearing assistance journey; the potential for improved understanding of the needs of different groups within the population; specific data analysis proposals. There were also four remaining responses about general aspects.

*Table 3. Free-text responses to the question "Dream big: what would you do with 'big audiological data'?", grouped to identified themes. Numbers in brackets indicate the number of statements per theme.*

| **Theme** - Topics | |
|---|---|
| **Predictors of outcomes** (5) | |
| | Comparison of outcomes in different dimensions; Related to HDs, socio-economic status; Enable a deeper look at cochlear implant outcomes; Understand what is a successful fitting (best time, variables important for HD selection and fitting); Perform comparative analysis of audiological data for different populations, service delivery models, and best practices to tailor to national and regional hearing health systems. |
| **Prevention and awareness** (7) | |
| | Discover trends; Public communication/knowledge transfer; Document costs and benefits for society; Quantify impact of hearing loss and hearing intervention; Increase access to hearing care; Early diagnosis; Causal relationships to other diseases. |
| **Personal hearing journey** (6) | |
| | Accompanying patients; Improved counselling of patients; Optimized, individual management plans for patients and clinicians; Development of personalized resources; Give control about their own hearing care; Find better ways to help patients hear better. |
| **Improved understanding of groups in the population** (9) | |
| | Needs of different population groups; Address groups that receive little attention today, including cultural and regional differences; Advance contextually-relevant interventions; Improve diagnostics and rehabilitation for rare cases; Study pathophysiology of specific phenotypes; Understand that individuals might not fit in any specific model; Answer population-representative questions; Create a worldwide map of hearing impairment; Contribute and do collaborative research. |
| **Analysis proposals** (10) | |
| | Unsupervised machine learning to detect patterns and associations; Personalized models of hearing loss; Agent-based audiological large language model; Decision-support for diagnostics or fitting of HDs; Clinical decision support system based on representative data from all continents; Audiological profiles, also related to, e.g., tinnitus, satisfaction, listening effort; Data-driven basis for early screening and intervention; Optimal combination of tests for better individualized and population-based diagnosis, understanding of demographic factors, risk factors; Enable a worldwide database with audiological tests and metadata for HDs; Informed gain prescription (first-fit) from all patients. |
| **Other responses** (4) | |
| | Advanced research will provide large benefits to the field; Novel research; Advancing research; Potential of predictive power while paying attention to include not only past data. |

When asked to identify, from a predefined list of categories, the two most important potential hurdles related to the envisioned large, global databases (Figure 3C), the most frequently selected option was the legal basis for data collection, i.e., matters related to data privacy, consent, or data ownership. This result was followed by — in descending order — the range of different types and conditions of audiological tests in use, data formats and structure, infrastructure, insufficient database documentation, business value for corporations potentially unclear, and data literacy. Three respondents specified other hurdles, such as how to achieve acceptance by clinicians, who would pay for such a repository, and whether sufficient data would be available.

*Survey Section 3: Data standards*

Table 4 shows the results to the respondents' understanding about data standards. The majority of respondents indicated that they were familiar with the concept of data standards and that there is a need for data standards in audiology. More than half of respondents were unaware of existing approaches to data standardisation, while slightly less than 30% had already contributed to the development of standards. Of those who had contributed to development, respondents indicated that this was for — in descending order — unaided and aided audiological test data, followed by “other” data, which they specified as data related to ototoxicity, cognition, and cochlear implantation—see Supplementary Material. In a follow-up question, respondents could describe known approaches and mentioned open standards (e.g., openEHR, OMOP), industrial standards (e.g., HIMSA’s Noah standard), and standards and guidelines established by societies and associations (American National Standards Institute, British Society of Audiology, American Speech-Language-Hearing Association, German Audiology Society). A majority of respondents declared an interest in following developments in this field, and overall 90% were interested or potentially interested in contributing to the development of standards.

*Table 4. Distribution of responses to survey questions related to data standards for audiology.*

| | | |
|---|---|---|
| **Are you familiar with the concept of data standards?**<br>[One answer possible.] | | |
| | Count | Percentage |
| Yes | 67 | 82% |
| No | 8 | 10% |
| Not sure | 7 | 9% |
| **Do we need data standards in audiology?**<br>[One answer possible.] | | |
| | Count | Percentage |
| Yes | 75 | 91% |
| No | 1 | 1% |
| Not sure | 6 | 7% |
| **Do you know existing standardisation approaches / attempts in audiology?**<br>[One answer possible.] | | |
| | Count | Percentage |
| Yes | 31 | 38% |
| No | 51 | 62% |
| **Do you work / contribute to a data standardisation initiative for audiology?**<br>[One answer possible.] | | |
| | Count | Percentage |
| Yes | 23 | 28% |
| No | 59 | 72% |
| **Are you interested in following the developments of data standards in audiology?**<br>[One answer possible.] | | |
| | Count | Percentage |
| Yes | 68 | 83% |
| No | 14 | 17% |
| **Are you interested in contributing to the development of data standards in audiology?**<br>[One answer possible.] | | |
| | Count | Percentage |
| Yes | 38 | 46% |
| No | 8 | 10% |
| Maybe, depending on how this would work in practice. | 36 | 44% |

When asked to identify, from a predefined list of categories, the two most important potential benefits of data standards, the most frequently selected option was comparing data from different locations (68%; 56 responses), followed by the opportunity to conduct research on real patient data, e.g., through electronic health records (50%; 41 responses), reproducibility through guidelines on reporting test conditions in publications and database documentation (34%; 28 responses), exchanging AI models and analysis tools, e.g., to apply them to different datasets (17%; 14 responses), building applications on databases (16%; 13 responses), and joint project applications (11%; 9 responses). When asked to identify, from a predefined list of categories, the two most important properties data standards should have to be successful, i.e., to be applicable, useful, and accepted, the most chosen property was flexibility, i.e., data standards should be applicable to many scenarios (48% of respondents; 39 responses), followed by intuitiveness and ease of understanding (38%; 31 responses), feasibility (30%; 25 responses), global adoption (23%; 19 responses), openness / availability (20%; 16 responses), community-driven (18%; 15 responses), defined by database experts (10%; 8 responses), and completeness (7%; 6 responses).

*Panel discussion*

**Overview**. The initial set of 74 survey responses were discussed during the VCCA2024's special session on big data and data standards in audiology with a panel of five experts. The panel discussion is summarised in the upcoming section as per the four main topics discussed: (1) the experts' motivations to participate in the panel; (2) state of the art of data standardisation in audiology; (3) challenges; and (4) recommendations.

**Panel members' motivation to participate in the panel discussion**. Panel members shared their motivations for participating in the discussion from different perspectives, i.e., as researchers, clinicians, database managers, or from combined perspectives. Specific motivations included the need and potential to reuse routinely collected clinical data and EHRs for research purposes, to improve clinical data quality, and to homogenise and integrate data from different locations and sources to enable big data analytics and AI applications. All of the experts had either worked on data standards or had expertise highly relevant for the development of data standards. The experts agreed with the survey results that the topic of data standards for audiology was important and timely and that audiology has lagged behind other medical fields in this regard.

**State of the art of data standardisation in audiology**. Overall, the discussion revealed the absence of a complete, comprehensive, and widely adopted data standard for the domain of audiology to date. OMOP and openEHR were mentioned as possible starting points, together with some of their key properties (e.g., being available for different health domains and pursuing a community-driven, flexible approach). The experts highlighted that knowledge of how to structure data is available but scattered across institutions. In addition, the data structures developed so far are not necessarily interoperable, i.e., do not follow a common format with approaches tailored to individual needs.

**Challenges**. In addition to challenges already identified as part of the community survey (see Figure 3), the experts referred to data challenges in the context of re-using clinical data and making it available for research purposes. In particular, poor quality of clinical data may prevent their re-use. It was suggested that the poor data quality may be related to, e.g., inconsistent entries in the same fields of a clinical software or incomplete documentation of test conditions. Data standards could help improve the structure and quality of data, which would benefit not only research but also clinical practice. However, it was also pointed out that meeting these higher data standards must not "interfere" with routine clinical work.

In addition, the panel emphasized the need to fully understand all the variations into which tests are used. For instance, different tests may be used for the same purpose, or the same tests may be performed under different conditions. Therefore, experts suggested that it would be imperative for data standard development in audiology to involve all relevant stakeholders, in order to capture many different perspectives and examples of current practice.

Another recurring theme in the discussion was whether the development of data standards for audiology should start within audiology or whether it would be beneficial to directly build a bridge to other medical domains. On the one hand, panel members emphasised that audiology-specific concepts, such as a common definition of pure-tone averages, should be addressed within the community. On the other hand, the panel stated that bridging to other medical domains is important to gain a better picture of the relationships between diseases and health conditions, and to continue to raise awareness of the importance of hearing loss to society.

**Recommendations**. The panel members emphasised that a data standard needs to encompass all possible data collections, their associated metadata and documentation, while at the same time being flexible enough to adapt to existing data standards and structures. They agreed that the most promising, feasible, and flexible approach would be to consider a range of standardisation approaches for audiology that institutions and laboratories could choose from. Crucially, however, these different standardisation approaches would need to allow for conversion between them. The panel also stressed that further work is needed to harmonise data collections across institutions. This step could be a part of the data standards implementation process. To achieve a plausible set of data standards, a balanced approach between the interests of the audiological community and an openness towards other medical fields would be required.

As a conclusion, the panel discussion confirmed the need and potential of data standards for audiology. Promising starting points exist but need to be further explored and developed, different stakeholders need to be involved, and an openness to other medical fields is required.

## Discussion

The objective of this study was to collect insights from the computational audiology community about the current status, needs, and potential of data standards for audiology, to discuss these insights with experts, and to derive how data standards for audiology could be achieved.

*Audiological data and potential for analysis of combined data*

The scope of audiological data collections documented by the survey encompassed a comprehensive range, including standard clinical audiometric tests, unaided and aided performance measures, HD specifications and settings, holistic predictors and outcome measures, as well as data collected through smartphone applications and wearable sensors. This diversity aligns with the wider literature on how big data sets have been used before (e.g., Dillard et al., 2020; Cantuaria et al., 2021; Sanchez-Lopez et al., 2021; Jafri et al, 2024; Laplante-Levesque et al., 2016). The potential applications for global audiological datasets, as identified by the survey, extended beyond those previously documented in literature (e.g., Lesica et al., 2021; Wasmann et al., 2021) and included HD outcome prediction, informed hearing loss prevention, participatory health care design, knowledge generation for rare hearing conditions, understanding cultural and regional differences in hearing health care, and numerous specific analysis protocols. These examples collectively demonstrate substantial community motivation for leveraging standardised data, and highlighted the heterogeneity of data that needs to be represented in comprehensive data standards for audiology.

*Community needs and familiarity with data standards*

The findings from our survey, panel discussion, and literature review emphasised the widespread interest in audiological data standardisation, with 91% of survey respondents confirming the necessity for standards in audiology. Despite this recognition, the current landscape revealed varying levels of engagement with standardisation processes. A mere 38% of respondents reported knowledge of existing standardisation approaches; 28% had actively contributed to data standards development to date. This gap between interest and active participation highlighted a significant opportunity for community information and engagement. Given that the sample of respondents likely mainly included computational audiologists brings up the question of how much lower the positive responses might have been if a wider audience had been included. It also underscores the importance of community information as a pressing first step to increase community involvement and engagement.

*Data standards in audiology*

Three standardisation approaches are suggested as viable frameworks and starting points for audiology: openEHR, OMOP, and the HIMSA's Noah standard. The openEHR approach offers flexibility to accommodate various data categories (e.g., diagnostic tests, medication/treatment) through its modular data architecture, consisting of archetypes and templates (Haarbrandt et al., 2018; Beale, 2002). While requiring technical knowledge to implement openEHR, its Clinical Knowledge Manager platform facilitates domain expert review (also see Table 1). The structured community review process for archetypes provides a clear pathway for achieving consensus on clinical concepts. OMOP offers adaptability through its extensible standardised vocabulary, which can accommodate diverse use cases. While primarily built for observational data, it has also been applied to clinical trial data (Orlova et al., 2020; Sung et al., 2024), a wide variety of data stored in the UK Biobank (Papez et al., 2022), and comprehensive patient data recorded by a variety of German local health systems including demographics, hospital visits, diagnoses, procedures, laboratory observations and medications, and other extension modules (Peng et al., 2023). Implementation, however, requires substantial effort and close collaboration between modelling and domain experts (Kohler et al., 2023). For an example of successful collaboration between a variety of stakeholders, see Sung et al. (2024). HIMSA's Noah standard, linked to HD and measurement device manufacturers, represents an industry de-facto domain standard with significant valuable knowledge incorporated. Noah can capture several audiological tests, especially typical clinical ones, as well as HD information. Contributions from the audiology community outside industry seem currently limited but possible, and connecting and integrating HIMSA's Noah standard with openEHR / OMOP would be beneficial to accommodate standardised audiological data from clinical, research, and industrial data sources.

A comprehensive data standard for audiology would ideally have the following key characteristics, as identified in the survey and panel discussion: flexibility for different scenarios, intuitive usability, feasible implementation, compatibility with clinical workflows, i.e., without disrupting routine data collections, and building a bridge to other medical fields. All three approaches could theoretically accommodate clinical and research data (of the different categories identified as important in the survey, see Figure 2A) without modifying the source data formats. openEHR and OMOP, in addition, offer substantial integration potential with other medical fields and systems, given their cross-domain implementation in health care, which is important in light of integrated care principles (WHO, 2019).

*Different stakeholders and community involvement*

By definition, standards are established and approved by a governing body (Benson & Grieve, 2021, Chapter 22). To date, there is no recognised body uniting all audiology stakeholders. Our survey aimed to engage the audiology community and put their voice at the centre of development. The survey primarily reached clinicians, researchers, and database experts, while previous studies by Laplante-Levesque et al. (2016) and Christensen et al. (2019) engaged industry partners and policymakers. The stakeholder composition proposed by Laplante-Levesque et al. (2016), i.e., hearing aid manufacturers, health care systems, professionals, and patients, generally aligned with our panel discussion recommendations, but omitted database experts and researchers — critical groups for ensuring technical feasibility and research applicability.

Community engagement depends crucially on individual effort and the overcoming of knowledge barriers. Our survey revealed that 46% of respondents expressed willingness to contribute to standardisation efforts. Another 44% indicated conditional willingness ("Maybe, depending on how this would work in practice"), suggesting that accessible contribution mechanisms and clear implementation guidance will be crucial for mobilising the audiological community. It is important to note that community involvement structures differ significantly across existing standards: openEHR provides formal review processes to any interested party for new archetypes; OMOP allows single researchers or groups to propose vocabulary extensions via SNOMED CT UK and its ATHENA platform for online publication of modelled concepts (also see Table 1), however, without requiring community consensus; HIMSA maintains proprietary control through its consortium. To overcome these limitations we suggest proof-of-concept studies between a small number of partners; these could serve as powerful motivators for broader participation.

*Conversion between standards*

Based on the panel discussion, parallel developments of different data standardisation formats for audiology seemed to be the preferred approach. To ensure this approach's success, it would be important to establish conversion mechanisms between them. This approach is common in medical informatics. Rinaldi & Thun (2021), for instance, successfully mapped microbiology data between openEHR, FHIR, and OMOP formats. While complete data and metadata descriptions are the main requirement for interoperability, conversion between standards allows to leverage specific advantages of each standard, for example, OMOP's optimisation for analysis (OHDSI, 2021) or openEHR's comprehensive metadata handling (Haarbrandt et al., 2018). The existing audiogram modelling in OMOP (Mehta et al., 2024) and the archetype development by the openEHR EFAS working group (EFAS Working Group, n.d.) represent valuable starting points for the parallel development strategy.

As many countries progress in health digitalisation and government initiatives begin to make health data available to clinicians, patients, and researchers through national portals, the development of audiology standards could follow the data standards used on these platforms. Examples include the French National Health Data System (SNDS) with their Health Data Hub (SNDS, n.d.), the German Portal for Medical Research Data (FDPG, FDPG, n.d), or the Danish Health Data Authority (DHDA; DHDA, n.d.). To the best of our knowledge, only the DHDA incorporates audiological data with its "Høremappen" (hearing folder), which is based on the Noah standard from HIMSA, the Cross-Enterprise Data Sharing (XDS) protocol, and additional metadata files. Aligning with such initiatives would allow for the integration of existing infrastructure, and facilitate further development. Conversion between different standards would allow national developments to be linked to global standards. However, audiological data models to represent all audiological data needs must first be developed before data can be stored in standardised formats.

*Implementation of data standards for audiology*

This study was an exploration of community interest towards audiological data standards. A detailed roadmap describing all steps until clinical implementation was beyond the scope of the study. However, we will briefly discuss the steps required to achieve standardisation. As mentioned above, the first step is a meaningful engagement of all stakeholders in developing data models. These stakeholders include researchers, clinicians, and data scientists from both the public and private sectors. Tools for modeling different standards and servers to store the data are also required. One example of an open-access tool for data standard modelling is the openEHR Archetype Designer (cf. Table 1). Before new data can be saved directly in agreed-upon standardised formats, the data models must be available and tested on a few datasets. This can be accomplished through proof-of-concept research studies. Only after this can clinical integration begin, for which the burden and cost related to implementation and maintenance should be estimated and reviewed. Following clinical integration, collecting data in a standardised format should not negatively affect clinical or experimental workflows because the format is defined in the data collection software, and the exact storage formats are not visible to users. Using a standardised format with defined metadata documentation will also improve the quality and consistency of the collected data.

Data standards have already been successfully implemented in other medical fields, such as cardiovascular disease and infection control. For example, the transformation to OMOP data models has been used to integrate data from two healthcare systems in the U.S., involving several hospitals, as well as catheter device and patient EHR data (Yu et al., 2023), or across four Danish health registries to correlate cardiovascular events with colorectal cancer (Zahid et al., 2025). The SmICS tool enables clinical decision-support for infection prevention and control based on openEHR data models (Biermann et al., 2025). In these studies, developing the necessary data models came before creating analyses and tools for the integrated data. Data standardization was essential for automatically assessing interoperability.

*Limitations of this study*

While the original survey was distributed via a variety of social media channels to a broad audiology community, its link to the VCCA made it likely that most respondents were researchers engaged in computational audiology. It is also possible that many of the respondents were more technically skilled and technology-minded than what could be expected from other stakeholder groups such as clinicians.

It is also important to acknowledge that answers may have been biased to the positive because we did not discuss or inquire about the potential burden of implementing community-agreed data standards. Working out the burden and costs for implementing, maintaining, and engaging with these standards has to be at the heart of their development if they are to be widely accepted.

*Additional challenges towards data exchange and big audiological data*

In addition to standardisation itself, other challenges must be addressed to enable data exchange and big data analysis in audiology. These challenges include navigating data privacy restrictions, ensuring data quality across collection methodologies (from digital databases to paper documentation), and developing appropriate infrastructures. While these aspects were beyond the scope of this study, they represent essential complementary considerations. Local collaborations with established data privacy protocols can serve as pilot projects before expanding to international scale. In the near future, the new European Health Data Space (EHDS) regulation (European Commission, 2025) should facilitate data sharing and pooling.

The combined aspects of data standards, data privacy, and robust infrastructure will ultimately determine the feasibility of international data exchange in audiology. However, data standards can facilitate exchange between institutions, even on a smaller collaborative scale. Ultimately, successful standardisation could also foster infrastructure development, by providing clear structural templates for database design, thereby advancing both standardisation and the technical capacity for data sharing across the field.

## Conclusions

This study evaluated the current status and potential of data standards for audiology, as an important prerequisite for large-scale data analysis, aimed at advancing research and patient care. The most important findings and implications are:

- The community survey revealed strong community support, a perceived need for data standards, and motivation to participate. However, familiarity with existing approaches was low, emphasising the need to collect and exchange more information about data standards.
- A comprehensive, community-accepted data standard for audiology requires further development of starting points. Available standards such as openEHR and OMOP have different advantages, and all of them could be leveraged by allowing conversion between standards. The latter is already common practice in other medical domains. Feasibility of integrating domain knowledge as incorporated in the HIMSA standard should be explored.
- Involvement of all relevant stakeholders (researchers, clinicians, industry, policy makers, database experts, and patients) is critical; first efforts will likely rely on a smaller group of contributors, while the broader community will be engaged to provide valuable feedback.
- Prototypical collaborative projects between two or more partners could help to achieve a proof of concept for the definition of data standards and the integration of databases, from which further development steps can be derived for future convergence towards international standards.
- Clear guidelines and tools are required to ensure high data quality without disrupting clinical workflows.

**Acknowledgements**

The authors would like to thank all members of the computational audiology community who filled in the survey, as well as the experts who joined the panel discussion, including Graham Naylor and Eugen Kludt. We would also like to thank Peter Kossek, Ruth Spriggs, and Antje Wulff for contributing their expertise in comparing different existing data standardisation initiatives. Finally, we would like to thank the organizers of the VCCA for giving us the opportunity to organise a special session on “Big data and data standards for audiology” at the VCCA2024.

**Declaration of Conflicting Interests**

C.V. and C.L. are employed by Sonova AG. The authors declare that this affiliation did not influence study design, data collection and analysis, decision to publish, or preparation of the manuscript.

**Data availability statement**

Data will be made available upon request.

**Funding**

M.B. was funded by the Deutsche Forschungsgemeinschaft (DFG, German Research Foundation) – Projektnummer 496819293, and supported by a grant from Fondation Pour l’Audition (FPA) to the CERIAH (FPA IDA10). This work has benefited from a French government grant managed by the Agence Nationale de la Recherche under the France 2030 programme, reference ANR-23-IAHU-0003. A.H. is supported by the National Institute for Health and Care Research (NIHR) Manchester Biomedical Research Centre (BRC) (NIHR203308).

Sonova AG provided funding in the form of salary for the authors C.V. and C.L., but did not have any additional role in the study design, data collection and analysis, decision to publish, or preparation of the manuscript. C.V. was also supported by the NIHR Manchester Biomedical Research Centre.

**Author contributions**

CV: Conceptualisation, survey design, data curation, formal analysis, investigation, methodology, software, writing--original draft preparation, writing--review and editing.

AH: Investigation, writing--review and editing.

CL: Investigation, software, writing--review and editing.

AP: Investigation, writing--review and editing.

JW: Investigation, writing--review and editing.

MB: Conceptualisation, funding, survey design, data curation, formal analysis, investigation, methodology, writing--original draft preparation, writing--review and editing.

All authors have read and agreed to the final version of the manuscript.

**Supplementary material**

An overview of all survey questions is provided below, with response options between brackets. Unless indicated otherwise, participants could only choose one response option.

Demographic questions:

1. Where are you originally from?
(Africa / Antarctica / Asia / Australia / Europe / North-America / South-America)

2. What is your current profession? [Multiple answers possible.]
(Audiologist / Data scientist / Educator / ENT doctor / Hearing aid acoustician / Informatician / Researcher / I prefer not to answer / Other, please specify)

3. How many years of professional experience do you have in the domain of audiology / hearing (in total)?
(0-1 / 2-3 / 4-5 / 6-10 / 11-20 / 21-30 / > 31 / I prefer not to answer.)

4. What is your study / educational background? [Multiple answers possible.]
(Artificial intelligence, machine learning / Audiology / Data science / Hearing aid acoustician school / Medicine / Natural sciences (math, physics, biology, neuroscience, ...) / Psychology / I prefer not to answer / Other, please specify)

5. Do you attend this conference for private / professional reasons? [Multiple answers possible.]
(Private / Professional / I prefer not to answer / Other, please specify)

Questions related to audiological data:

1. Which data are in your opinion important in audiology (relevant for any purpose you can imagine)? [Multiple answers possible.]
(Audiological test data (unaided) / Audiological test data (aided) / Demographic and anamnesis data / Data logging of hearing devices / Data from hearables, wearables, consumer electronic devices / Hearing device fitting data / Hearing screening data / Self-reported benefit and satisfaction data / Other health data / Other, please specify)

2. Do you work with audiological data (in general)? [Multiple answers possible.]
(Yes, I collect data (as part of clinical practice) / Yes, I collect data (as part of research studies) / Yes, I interpret data (for individual patients / participants) / Yes, I analyse data (for groups of patients / subjects) / No, I don’t work with data / Other, please specify)

3. Which of the above-mentioned activities (working with audiological data) is most frequent in your work routine?
(Collecting data (as part of clinical practice) / Collecting data (as part of research studies) / Interpreting data (for individual patients / subjects) / Analyzing data (for groups of patients / subjects) / All activities are similarly frequent / Not applicable (since I don’t work with data) / Other, please specify)

Detailed questions related to audiological data with the following additional instructions: "*Answer the questions in this section for your most frequent data activity as indicated before (collection / interpretation (individual) / analysis (group))*":

1. Which data do you collect / interpret / analyse? [Multiple answers possible.] (Demographic and anamnesis data / Audiological test data (unaided) / Audiological test data (aided) / Hearing device fitting data / Data logging of hearing devices / Data from wearables / Data from hearables / Data from over-the-counter (OTC) hearing aids / Hearing screening data / Self-reported benefit and satisfaction data / Other health data / Other, please specify)

2. Which data do you collect / interpret / analyse - demographic and anamnesis data. [Multiple answers possible.]
   (Age / Gender / Anamnesis data related to hearing loss / Anamnesis data related to comorbidities in its broadest sense (vision loss, dexterity, cognition, tinnitus...) / Health literacy / Social support / Technological savviness / Socio-economic status / Motivation / Validated questionnaires / Homemade questionnaires / Hearing screening data / Not applicable / Other, please specify)

3. Which data do you collect / interpret / analyse - diagnostic tests (unaided). [Multiple answers possible.]
   (Audiogram (air conduction) / Audiogram (bone conduction) / Speech intelligibility in quiet / Speech intelligibility in noise / Otoscopy / Tympanometry / Spectral resolution / Temporal resolution / Loudness scaling / Localisation / Vestibular tests / Otoacoustic emission / Acoustic reflex / Auditory brainstem response / EEG / Non-auditory holistic diagnostic tests / Not applicable / Other, please specify)

4. Which data do you collect / interpret / analyse - diagnostic tests (aided). [Multiple answers possible.]
   (Audiogram (air conduction) / Speech intelligibility in quiet / Speech intelligibility in noise / Real ear measures / Real ear to coupler differences / Not applicable / Other, please specify)

5. Which data do you collect / interpret / analyse - hearing device fitting data (hearing aid) and data logging [Multiple answers possible.]
   (Gain / Compression ratio / Maximum power output / Device information / Signal processing / Fitting formula / Wearing time / Wearing time in different hearing aid programmes / Verification (fit to target) / Not applicable / Other, please specify)

6. Which data do you collect / interpret / analyse - hearing device fitting data (cochlear implant). [Multiple answers possible.]
   (T-levels / M-levels / Number of electrodes / Insertion depth / Impedance data / Stimulation rate / Coding strategy / Device information / Not applicable / Other, please specify)

7. Which data do you collect / interpret / analyse - data from wearables. [Open ended question.]

8. Which data do you collect / interpret / analyse - data from hearables. [Open ended question.]

9. Which data do you collect / interpret / analyse - data from over-the-counter (OTC) hearing aids. [Open ended question.]

10. Which data do you collect / interpret / analyse - hearing screening data. [Open ended question.]

11. Which data do you collect / interpret / analyse - self-reported benefit and satisfaction data. [Open ended question.]

12. Which data do you collect / interpret / analyse - other health data. [Open ended question.]

13. In which format do you store your data? [Multiple answers possible.]
    (One global well-structured database (integrating data measured with different devices / Multiple well-structured databases / Digital files from different measurement devices / Paper files / Digital copies of paper files / Data manually entered in a digital spreadsheet / Not applicable / Other, please specify )

14. Which software do you use for data collection and management? [Open ended question.]

15. Do you encounter any difficulties in using your data for their intended purpose? [Multiple answers possible.]
    (No, smooth sailing / Yes, infrastructure problems (e.g., multiple databases etc.) / Yes, legal problems (e.g., consent, data sharing, etc.) / Yes, data analysis, statistical problems / Yes, no resources to analyse data / Other, please specify)

Questions related to big audiological data with the following additional instructions: "*To date, a large variety of data is collected in clinical as well as research settings, including studies with large numbers of patients. These data are typically collected in local databases, exhibiting different data structures, choice of tests, or varying data quality*.":

1. Imagine that all audiological data worldwide would be available in one database: what could be the benefit from analyses on this database? [Choose the two most important options for you.]
   (Advancing research / Advancing patient care / Preventative health care / Training Artificial Intelligence (AI) models / Digital twins / Epidemiological, population-representative studies / Collecting data of parts of the world that we currently get little data from (e.g., to investigate cultural differences, etc.) / I prefer not to answer / Other, please specify)

2. For which purpose could AI models be relevant? [Choose the two most important options for you.]
   (Decision-support / Data-driven characterisation of the population / Definition of optimal test batteries for patient characterisation / Optimisation of hearing device algorithms / Knowledge generation / I prefer not to answer / Other, please specify)

3. Who could benefit from analyses on this database? [Multiple answers possible.]
   (Clinicians / Researchers / Patients / Health policy-makers / Health insurers / I don't see any potential benefit / I prefer not to answer / Other, please specify)

4. What are the biggest hurdles that you see? [Choose the two most important options for you.]
   (Legal basis for data collection and matters related to data privacy, such as consent, data ownership, … / Infrastructure / Data formats and structure / Different choice and conditions of audiological tests in general / Insufficient database documentation / Business value for corporations might be unclear / Data literacy / I don't see any hurdles / I prefer not to answer / Other, please specify)

5. Dream big: what would you do with "big audiological data" (all data worldwide accessible in one database, in a common format)? [Open ended question.]

Questions related to data standards:

1. Are you familiar with the concept of data standards?
   (Yes / No / Not sure)

2. Do we need data standards in audiology?
   (Yes / No / Not sure)

3. Do we need data standards in audiology? - Optional comment. [Open ended question.]

4. Which potential benefit of data standards do you see? (Choose the two most important options for you.)
   (Linking or comparing data from different locations, e.g., clinics, research institutes, countries, … / Performing research on real patient data, e.g., from electronic health records / Joint project applications / Guidelines on reporting test conditions etc., in publications and database documentation, to support reproducibility / Building applications on databases / Exchanging AI models and analysis tools, to apply them on different data sets / I don't see any potential benefit / I prefer not to answer / Other, please specify)

5. Which properties should a successful data standard have (to be applicable, useful, accepted, …)? [Choose the two most important options for you.]
   (Intuitive, easy to understand / Defined by the community / Defined by database experts / Applicable to many scenarios, flexible / Complete / Feasible implementation / Open tools available for definition of data standards / Global adoption of the standard / I prefer not to answer / Other, please specify)

6. Do you know existing standardisation approaches / attempts in the field of audiology? (Yes / No)

7. If yes, which? [Open ended question.]

8. Do you work / contribute to a standardisation initiative for audiology? (Yes / No)

9. If yes, which? [Open ended question.]

10. If yes, for which type of data? [Multiple answers possible.] (Audiological test data (unaided) / Audiological test data (aided) / Data logging of hearing devices / Data from wearables / Data from hearables / Data from over-the-counter (OTC) hearing aids / Demographic and anamnesis data / Hearing device fitting data / Hearing screening data / Self-reported benefit and satisfaction data / Other health data / Other, please specify)

Closing questions:

1. Do you already have a question for the panellists to be asked during the session? Please state your question here: [Open ended question.]

2. Are you interested in contributing to the development of data standards in audiology? (Yes / No / Maybe, depending on how this would work in practice.)

3. Do you consent to the fact that the data collected through this survey would be analysed and used for a scientific publication? (Yes / No)

An overview of responses to survey questions that were not reported or discussed in the manuscript. Unless indicated otherwise, participants could only choose one response option.

**Demographic questions:**

Table 1. Responses to survey question “What is your study / educational background?” [Multiple answers possible.]

| **What is your study / educational background?** [Multiple answers possible.] | | |
|---|---|---|
| | Count | Percentage |
| Audiology | 38 | 46% |
| Natural sciences (mathematics,physics, biology, neurosciences, ...) | 22 | 27% |
| Artificial intelligence / machine learning | 14 | 17% |
| Data science | 11 | 13% |
| Psychology | 9 | 11% |
| Medicine | 7 | 9% |
| Hearing aid acoustician school | 3 | 4% |
| Prefer not to answer | 0 | 0% |
| Other | 15 | 18% |

**Detailed questions related to audiological data with the following additional instructions:**
"*Answer the questions in this section for your most frequent data activity as indicated before (collection / interpretation (individual) / analysis (group))*":

Table 2. Responses to survey question "Which data do you collect / interpret / analyse?" [Multiple answers possible.]

| **Which data do you collect / interpret / analyse?** [Multiple answers possible.] | | |
|---|---|---|
| | Count | Percentage |
| Audiological test data (unaided) | 59 | 72% |
| Demographic and anamnesis data | 50 | 61% |
| Audiological test data (aided) | 49 | 60% |
| Self-reported benefit and satisfaction data | 46 | 56% |
| Data logging of hearing devices | 29 | 35% |
| Hearing device fitting data | 29 | 35% |
| Hearing screening data | 26 | 32% |
| Other health data | 21 | 26% |
| Data from wearables | 8 | 10% |
| NA | 7 | 9% |
| Data from hearables | 6 | 7% |
| Data from over-the-counter (OTC) hearing aids | 6 | 7% |
| Other | 6 | 7% |

Table 3. Responses to survey question “Which data do you collect / interpret / analyse, specifically related to demographic and anamnesis data?” [Multiple answers possible.]

| **Which data do you collect / interpret / analyse, specifically related to demographic and anamnesis data?** [Multiple answers possible.] | | |
|---|---|---|
| | Count | Percentage |
| Age | 69 | 84% |
| Gender | 65 | 79% |
| Anamnesis data related to hearing loss | 51 | 62% |
| Validated questionnaires | 48 | 59% |
| Anamnesis data related to comorbidities in its broadest sense (vision loss, dexterity, cognition, tinnitus… | 37 | 45% |
| Hearing screening data | 32 | 39% |
| Homemade questionnaires | 29 | 35% |
| Socio-economic status | 24 | 29% |
| Technological savviness | 16 | 20% |
| Motivation | 14 | 17% |
| Health literacy | 11 | 13% |
| Social support | 8 | 10% |
| Not applicable | 8 | 10% |
| Other | 6 | 7% |

Table 4. Responses to survey question “Which data do you collect / interpret / analyse, specifically related to diagnostic tests (unaided)?” [Multiple answers possible.]

| **Which data do you collect / interpret / analyse, specifically related to diagnostic tests (unaided)?** [Multiple answers possible.] | | |
|---|---|---|
| | Count | Percentage |
| Audiogram (air conduction) | 67 | 82% |
| Speech intelligibility in noise | 60 | 73% |
| Speech intelligibility in quiet | 49 | 60% |
| Audiogram (bone conduction) | 47 | 57% |
| Tympanometry | 35 | 43% |
| Auditory brainstem response | 30 | 37% |
| Otoacoustic emission | 30 | 37% |
| Otoscopy | 30 | 37% |
| Acoustic reflex | 29 | 35% |
| Loudness scaling | 20 | 24% |
| Localisation | 19 | 23% |
| Spectral resolution | 19 | 23% |
| EEG | 16 | 20% |
| Temporal resolution | 16 | 20% |
| Vestibular tests | 11 | 13% |
| Not applicable | 10 | 12% |
| Other | 9 | 11% |
| Non-auditory holistic diagnostic tests | 7 | 9% |

Table 5. Responses to survey question “Which data do you collect / interpret / analyse, specifically related to diagnostic tests (aided)? [Multiple answers possible.]

| **Which data do you collect / interpret / analyse, specifically related to diagnostic tests (aided)?** [Multiple answers possible.] | | |
|---|---|---|
| | Count | Percentage |
| Speech intelligibility in noise | 57 | 70% |
| Audiogram (air conduction) | 52 | 63% |
| Speech intelligibility in quiet | 52 | 63% |
| Real ear measures | 23 | 28% |
| Not applicable | 16 | 20% |
| Real ear to coupler differences | 9 | 11% |
| Other | 3 | 4% |

Table 6. Responses to survey question “Which data do you collect / interpret / analyse, specifically related to hearing device fitting data and (hearing aid) data logging? [Multiple answers possible.]

| **Which data do you collect / interpret / analyse, specifically related to hearing device fitting data and (hearing aid) data logging?** [Multiple answers possible.] | | |
|---|---|---|
| | Count | Percentage |
| Not applicable | 37 | 45% |
| Gain | 33 | 40% |
| Wearing time | 31 | 38% |
| Device information | 30 | 37% |
| Fitting formula | 24 | 29% |
| Verification (fit to target) | 22 | 27% |
| Wearing time in different hearing aid programmes | 20 | 24% |
| Compression ratio | 19 | 23% |
| Signal processing | 19 | 23% |
| Maximum power output | 17 | 21% |
| Other | 2 | 2% |

Table 7. Responses to survey question "Which data do you collect / interpret / analyse, specifically related to hearing device fitting data (cochlear implant)? [Multiple answers possible.]

| **Which data do you collect / interpret / analyse, specifically related to hearing device fitting data (cochlear implant)?** [Multiple answers possible.] | | |
|---|---|---|
| | Count | Percentage |
| Not applicable | 57 | 70% |
| Device information | 21 | 26% |
| Number of electrodes | 17 | 21% |
| Coding strategy | 16 | 20% |
| T-levels | 16 | 20% |
| Impedance data | 14 | 17% |
| M-levels | 14 | 17% |
| Stimulation rate | 14 | 17% |
| Insertion depth | 11 | 13% |
| Other | 2 | 2% |

Which data do you collect / interpret / analyse - data from wearables. [Open ended question.]

Which data do you collect / interpret / analyse - data from hearables. [Open ended question.]

Which data do you collect / interpret / analyse - data from over-the-counter (OTC) hearing aids. [Open ended question.]

Which data do you collect / interpret / analyse - hearing screening data. [Open ended question.]

Which data do you collect / interpret / analyse - self-reported benefit and satisfaction data. [Open ended question.]

Which data do you collect / interpret / analyse - other health data. [Open ended question.]

Which software do you use for data collection and management? [Open ended question.]

**Questions related to big audiological data with the following additional instructions**:
"*To date, a large variety of data is collected in clinical as well as research settings, including studies with large numbers of patients. These data are typically collected in local databases, exhibiting different data structures, choice of tests, or varying data quality*.":

Table 8. Responses to survey question "For which purpose could AI models be relevant?" [Choose the two most important options for you.]

| **For which purpose could AI models be relevant?**<br>[Choose the two most important options for you.] | | |
|---|---|---|
| | Count | Percentage |
| Decision-support | 42 | 51% |
| Data-driven characterisation of the population | 39 | 48% |
| Optimisation of hearing device algorithms | 35 | 43% |
| Definition of optimal test batteries for patient characterisation | 22 | 27% |
| Knowledge generation | 16 | 20% |
| I prefer not to answer | 2 | 2% |

**Questions related to data standards**:
Table 9. Responses to survey question "Do you work / contribute to a standardisation initiative for audiology? If yes, for which type of data?" [Multiple answers possible.]

| **Do you work / contribute to a standardisation initiative for audiology? If yes, for which type of data?**<br>[Multiple answers possible.] | | |
|---|---|---|
| | Count | Percentage |
| Not applicable | 56 | 68% |
| Audiological test data (unaided) | 14 | 17% |
| Audiological test data (aided) | 9 | 11% |
| Other | 6 | 7% |
| Demographic and anamnesis data | 5 | 6% |
| Other health data | 5 | 6% |
| Data logging of hearing devices | 4 | 5% |
| Self-reported benefit and satisfaction data | 4 | 5% |
| Hearing screening data | 3 | 4% |
| Hearing device fitting data | 1 | 1% |